\documentclass[fleqn,usenatbib]{mnras}
\usepackage{graphicx}
\usepackage{natbib}
\usepackage{txfonts}
\usepackage{lastpage}
\usepackage{tikz}
\usepackage{graphics,wrapfig,times}
\usepackage[absolute]{textpos}

\begin{document}

 \title[Are there some eccentric binaries below 1 day?]{Are there really some eccentric eclipsing binaries having orbital periods\\ below 1 day?}

 \author[Zasche \& Wolf]{Zasche,~P. \thanks{E-mail: petr.zasche@matfyz.cuni.cz}  \& Wolf, M. \\
 Charles University, Faculty of Mathematics and Physics, Astronomical Institute, V~Hole\v{s}ovi\v{c}k\'ach 2, CZ-180~00, Praha 8, Czech Republic}

\date{\today}
\pagerange{\pageref{firstpage}--\pageref{LastPage}} \pubyear{2025} \maketitle \label{firstpage}

\begin{abstract}
The stellar binary systems are classical tools for studying dynamics, and tidal circularization is
one of the processes taking place in these. During the last few years several dozens of systems
have been published supposedly having orbital periods shorter than 1~day, and were also published
having eccentric orbits. Such a configuration is due to their spatial closeness, and the tides
very improbable. Therefore, we performed a thorough analysis of all these eclipsing binaries
(>140) showing that the vast majority of the published systems ($>98\%$) are in fact circular.
Especially, the eclipsing binaries identified and published from the GAIA~DR3 catalogue suffer
from a lack of data, and about 50\% had incorrectly derived periods. This mostly incorrect
identification of the eccentric orbits is due to automatic pipelines and routines running without
human checking/validating, and also due to the large number of systems with (evolving) spots on
the surfaces of either of the stars. Only two eccentric orbits were found below the 1-day period
limit, while the star TIC~1190662 with orbital period of about 0.888~days have surprisingly large
value of eccentricity about 0.21.
 \medskip
\end{abstract}



\section{Introduction and motivation}

Classical eclipsing binaries are often considered the cornerstones of modern stellar astrophysics.
Their use for the calibration of theories and models has been discussed in many papers and
monographs (see, e.g. \citealt{2010A&ARv..18...67T}). This is also true for the tidal
circularization theories (see e.g. \citealt{1977A&A....57..383Z}, \citealt{1997A&A...318..187C},
or \citealt{1992ApJ...395..259T}). During the last decades, a huge effort has been made to
identify close binaries and to construct so-called P-e diagrams, where such circularization can be
clearly seen (see, e.g. \citealt{2008EAS....29....1M}, \citealt{2021ApJ...921..117T},
\citealt{2024A&A...691A.242I}).

However, one can ask an apparently simple question: are there any proven eccentric binary systems
below the orbital period of 1~day? As we show below, the answer is not as straightforward.
Answering this question should obviously provide us with very useful hints for testing the
circularization theories due to the sharp limit of the period cut-off for eccentric orbits. This
circularization period, below which only circular orbits are found, is still a matter of intensive
discussion but probably lies between 1 and 10~days, however, it also depends on the type of star.
For hotter stars, shorter periods with eccentric orbits also exist, while for cooler stars, longer
periods are already circularised (see e.g \citealt{2024A&A...691A.242I}). Therefore, finding a
definitively proven eccentric binary with a period below the 1 day limit would be important for
theoreticians. However, such a task is quite challenging, as can be seen in the following.

\begin{table*}
  \caption{Summarization of the systems.}  \label{systemsInfo}
  \scalebox{0.77}{
  \begin{tabular}{c c c c| c c c | c c c }\\[-3mm]
\hline \hline\\[-3mm]
    \multicolumn{4}{c|}{Identification}&   \multicolumn{3}{c|}{Original information}&  \multicolumn{3}{c}{Our verdict}    \\
 Target name                           &  TESS number  & RA [J2000.0]& DE [J2000.0] &  $P$ [d]  &   $e$ &    Reference           &  $P$ [d]   &  $e$  & comment/remark    \\  \hline\noalign{\smallskip}
 Gaia DR3 535964265885838720  & TIC 351611000 & 00 52 28.53 & +69 53 05.37 & 0.977425 & 0.0187 & \cite{Oddo2026} & 0.977426 & 0.00 & incorrect identification \\ 
 Gaia DR3 452750099161604736  & TIC 302233773 & 02 45 00.92 & +52 39 21.65 & 0.310142 & 0.0054 & \cite{Oddo2026} & 0.310072 & 0.00 & incorrect identification \\  
 Gaia DR3 3252605057117991552 & TIC 250131338 & 04 12 33.24 & -02 06 16.26 & 0.660856 & 0.0773 & \cite{Oddo2026} & 0.660856 & 0.00 & detached, spot \\ %
 Gaia DR3 2902056526567248384 & TIC 140211568 & 05 39 07.27 & -32 06 23.22 & 0.274053 & 0.0280 & \cite{Oddo2026} & 0.274054 & 0.00 & detached, spot \\ 
 Gaia DR3 3000683922169783296 & TIC 34311799  & 06 19 27.31 & -12 22 39.28 & 0.614671 & 0.0408 & \cite{Oddo2026} & 0.614666 & 0.00 & incorrect identification, spot, flares \\  
 Gaia DR3 940135463122345088  & TIC 68388891  & 06 51 26.01 & +34 28 04.69 & 0.534059 & 0.0208 & \cite{Oddo2026} & 0.534062 & 0.00 & detached, evolving spot \\ 
 Gaia DR3 5565496073306796800 & TIC 157345367 & 07 04 11.99 & -38 52 15.48 & 0.478058 & 0.0485 & \cite{Oddo2026} & --       &      & not eclipsing binary at all \\ 
 IO Cha                       & TIC 454366433 & 11 13 33.56 & -76 35 37.47 & 0.950716 & 0.0701 & \cite{Oddo2026} & 0.950710 & 0.00 & detached, flares \\  
 Gaia DR3 5376567883659319040 & TIC 162207682 & 11 15 07.30 & -44 46 48.09 & 0.808202 & 0.0563 & \cite{Oddo2026} & 0.808206 & 0.00 & incorrect identification \\ 
 ATO J172.0677-26.2415        & TIC 452698353 & 11 28 16.29 & -26 14 29.48 & 0.484362 & 0.0780 & \cite{Oddo2026} & 0.484359 & 0.00 & detached, evolving spot \\ 
 Gaia DR3 5860580289609692032 & TIC 309679390 & 12 19 14.10 & -65 57 28.51 & 0.859538 & 0.0212 & \cite{Oddo2026} & 0.85954  & 0.00 & incorrect identification \\ 
 Gaia DR3 3956695504263812224 & TIC 198485676 & 12 53 26.23 & +22 47 00.86 & 0.954624 & 0.0770 & \cite{Oddo2026} & 0.954625 & 0.00 & detached, spot \\ 
 Gaia DR3 6082873812379516032 & TIC 241413538 & 13 40 03.64 & -48 07 28.35 & 0.716733 & 0.1474 & \cite{Oddo2026} & 0.716385 & 0.00 & detached, evolving spot \\
 Gaia DR3 6104539001808460928 & TIC 241841997 & 14 27 58.42 & -39 23 28.25 & 0.927313 & 0.1901 & \cite{Oddo2026} &   --     &      & not eclipsing binary at all \\ 
 Gaia DR3 5872857230260862976 & TIC 454233941 & 15 03 44.24 & -65 20 28.11 & 0.706837 & 0.1693 & \cite{Oddo2026} & 0.706835 & 0.00 & asymmetric LC, spots \\
 ASAS J160556-3737.6          & TIC 69874547  & 16 05 56.04 & -37 37 47.76 & 0.794972 & 0.0091 & \cite{Oddo2026} & 0.794975 & 0.00 & evolving spots, flares \\  
 Gaia DR3 1431657636508961536 & TIC 199611396 & 16 37 02.18 & +57 56 43.59 & 0.525905 & 0.1071 & \cite{Oddo2026} & 0.525906 & 0.00 & detached, evolving spot, flares\\ 
 CRTS J174534.5+364417        & TIC 116020737 & 17 45 34.57 & +36 44 17.59 & 0.351089 & 0.0028 & \cite{Oddo2026} & 0.351087 & 0.00 & detached, spot \\ 
 Gaia DR3 4523378711985803392 & TIC 157669515 & 18 20 53.12 & +17 27 21.75 & 0.866844 & 0.0445 & \cite{Oddo2026} & 0.866850 & 0.01 & incorrect identification\\ 
 CRTS J194207.0-475341        & TIC 166282532 & 19 42 07.09 & -47 53 41.47 & 0.478491 & 0.0045 & \cite{Oddo2026} & 0.478490 & 0.00 & detached, evolving spot \\  
 CzeV1379                     & TIC 256362692 & 20 01 10.32 & +23 04 16.49 & 0.668858 & 0.0909 & \cite{Oddo2026} & 0.668859 & 0.00 & detached, spot, flares \\  
 WISE J211609.7+485954        & TIC 62602104  & 21 16 09.73 & +48 59 54.32 & 0.421117 & 0.0116 & \cite{Oddo2026} & 0.421199 & 0.00 & detached, no spot \\ 
 ATO J325.2040+41.8781        & TIC 419664745 & 21 40 48.98 & +41 52 41.62 & 0.843716 & 0.0382 & \cite{Oddo2026} & 0.843698 & 0.00 & detached, evolving spot \\ 
 Gaia DR3 2198042564210216448 & TIC 469674899 & 21 57 03.94 & +55 13 10.23 & 0.853245 & 0.0026 & \cite{Oddo2026} & 0.853260 & 0.00 & incorrect identification \\ 
 Gaia DR3 2589992273084612480 & TIC 58844298  & 01 18 37.91 & +13 16 27.65 & 0.281861 & 0.1539 & \cite{Gaia}                & 1.325294 &      & incorrect Gaia period \\
 Gaia DR3 4750134327870521728 & TIC 153066866 & 03 17 17.84 & -47 09 02.26 & 0.304962 & 0.1231 & \cite{Gaia}                & 0.304961 & 0.00 & circular, contact \\ 
 Gaia DR3 4803450226520223360 & TIC 151562520 & 05 51 14.69 & -43 32 24.96 & 0.936887 & 0.7255 & \cite{Gaia}                & 0.93688  & 0.00 & circular, contact \\ 
 Gaia DR3 4893790057525394816 & TIC 594833    & 04 48 02.30 & -25 10 15.99 & 0.619633 & 0.1340 & \cite{Gaia}                & 0.619631 & 0.00 & detached, circular \\ 
 Gaia DR3 4912189937939000448 & TIC 229155767 & 01 54 22.95 & -53 44 57.26 & 0.656211 & 0.1476 & \cite{Gaia}                & 0.656213 & 0.00 & circular, detached, no spots \\ 
 Gaia DR3 4938677551009080192 & TIC 166678085 & 02 20 52.62 & -48 56 40.33 & 0.596795 & 0.5710 & \cite{Gaia}                & 1.193577 &      & incorrect Gaia period \\ 
 Gaia DR3 5050809006688429056 & TIC 209405585 & 02 58 59.46 & -35 23 42.88 & 0.407424 & 0.1818 & \cite{Gaia}                & 0.407423 & 0.00 & circular, contact \\ 
 Gaia DR3 5224524112937179136 & TIC 454608316 & 11 31 25.55 & -76 39 34.26 & 0.651153 & 0.8360 & \cite{Gaia}                & 0.65116  & 0.00 & circular, contact \\ 
 Gaia DR3 5228646392496624000 & TIC 401597427 & 11 13 20.01 & -70 01 30.00 & 0.221498 & 0.2756 & \cite{Gaia}                & 3.749458 &      & incorrect Gaia period \\ 
 Gaia DR3 5235373273369227136 & TIC 294826848 & 11 18 36.20 & -67 20 26.71 & 0.213624 & 0.7147 & \cite{Gaia}                & 6.083029 &      & incorrect Gaia period \\ 
 Gaia DR3 5237175437324639360 & TIC 290889718 & 11 35 41.40 & -65 36 46.51 & 0.475600 & 0.6455 & \cite{Gaia}                & 3.448625 &      & incorrect Gaia period \\ 
 Gaia DR3 5241513392990746752 & TIC 464968980 & 10 48 36.48 & -62 58 33.89 & 0.577084 & 0.1231 & \cite{Gaia}                & 0.577083 & 0.00 & circular, contact \\ 
 Gaia DR3 5328332514714980352 & TIC 400084343 & 08 54 41.71 & -49 21 39.64 & 0.722764 & 0.8950 & \cite{Gaia}                & 3.17525  &      & incorrect Gaia period \\  
 Gaia DR3 5333720877610505600 & TIC 316795958 & 11 28 28.01 & -63 20 44.94 & 0.738181 & 0.5529 & \cite{Gaia}                & 1.476339 &      & incorrect Gaia period \\
 Gaia DR3 5338466571641700608 & TIC 466674418 & 11 05 54.22 & -60 05 20.45 & 0.727081 & 0.2620 & \cite{Gaia}                & 0.727078 & 0.00 & asymmetric LC, spot \\
 Gaia DR3 5345008051667423616 & TIC 452039454 & 11 33 38.14 & -54 28 29.50 & 0.238144 & 0.9635 & \cite{Gaia}                & 1.88765  &      & incorrect Gaia period \\
 Gaia DR3 5351271999459508224 & TIC 459062950 & 10 46 26.11 & -57 05 33.71 & 0.960418 & 0.7223 & \cite{Gaia}                & 4.84552  &      & incorrect Gaia period \\  
 Gaia DR3 5374077799061184384 & TIC 82768717  & 11 19 43.30 & -48 26 21.33 & 0.526593 & 0.4621 & \cite{Gaia}                & 4.74446  &      & incorrect Gaia period \\
 Gaia DR3 5404658550339657344 & TIC 132297733 & 09 54 20.77 & -52 51 11.53 & 0.924130 & 0.4265 & \cite{Gaia}                & 0.92413  & 0.00 & circular, contact \\ 
 Gaia DR3 555757571125248256  & TIC 260366084 & 03 48 40.86 & +79 21 57.72 & 0.335179 & 0.0853 & \cite{Gaia}                & 0.335178 & 0.00 & circular, contact \\ 
 Gaia DR3 5565790089587580160 & TIC 157344363 & 07 04 09.18 & -38 01 22.63 & 0.515003 & 0.8968 & \cite{Gaia}                & 9.88439  &      & incorrect Gaia period \\ 
 Gaia DR3 5584496733747162368 & TIC 173637300 & 07 30 00.76 & -41 09 42.17 & 0.656571 & 0.4105 & \cite{Gaia}                & 0.656578 & 0.00 & circular, detached, spot \\ 
 Gaia DR3 5694696626675122944 & TIC 155953553 & 08 16 27.76 & -25 33 23.27 & 0.921298 & 0.6845 & \cite{Gaia}                & 1.842565 &      & incorrect Gaia period \\ 
 Gaia DR3 5707658425651621120 & TIC 143525810 & 08 22 11.60 & -18 51 05.05 & 0.543804 & 0.4532 & \cite{Gaia}                & 1.575053 &      & incorrect Gaia period \\ 
 Gaia DR3 5712304991851559040 & TIC 142425464 & 07 54 32.20 & -21 18 26.52 & 0.245075 & 0.5353 & \cite{Gaia}                & 11.9955  &      & incorrect Gaia period \\
 Gaia DR3 5722679816070129280 & TIC 386313950 & 08 30 17.84 & -14 45 20.76 & 0.591561 & 0.2577 & \cite{Gaia}                & 0.591562 & 0.00 & circular, contact \\ 
 Gaia DR3 580107729293714176  & TIC 270678757 & 09 07 36.02 & +05 06 02.03 & 0.810569 & 0.8128 & \cite{Gaia}                & 0.810571 & 0.00 & circular, contact \\  
 Gaia DR3 5818991811993378816 & TIC 427014649 & 15 50 17.67 & -70 50 50.68 & 0.566636 & 0.5558 & \cite{Gaia}                & 0.566638 & 0.00 & circular, contact \\ 
 Gaia DR3 5863204617671248128 & TIC 438770125 & 13 01 47.89 & -62 46 21.51 & 0.433890 & 0.7718 & \cite{Gaia}                &  --      &      & not eclipsing binary at all \\ 
 Gaia DR3 5905395073532042240 & TIC 1154831998& 14 42 26.19 & -48 53 48.03 & 0.752919 & 0.4160 & \cite{Gaia}                & 0.752926 & 0.00 & circular, spot \\ 
 Gaia DR3 5977220812410453888 & TIC 41626231  & 17 05 50.86 & -35 30 02.05 & 0.306942 & 0.5195 & \cite{Gaia}                & 4.57101  &      & incorrect Gaia period \\ 
 Gaia DR3 5981170464289389056 & TIC 286735720 & 15 54 15.21 & -52 35 59.07 & 0.374669 & 0.5711 & \cite{Gaia}                & 3.060480 &      & incorrect Gaia period \\ 
 Gaia DR3 6010137647876035712 & TIC 301047011 & 15 55 06.35 & -37 57 20.47 & 0.448695 & 1.0263 & \cite{Gaia}                & 3.589812 &      & incorrect Gaia period \\ 
 Gaia DR3 6134365999190807424 & TIC 248206606 & 12 58 28.14 & -45 57 59.31 & 0.939124 & 0.5824 & \cite{Gaia}                & 26.78965 &      & incorrect Gaia period \\ 
 Gaia DR3 6139021400142398848 & TIC 359628375 & 13 02 11.60 & -42 29 12.99 & 0.691793 & 0.4645 & \cite{Gaia}                & 0.691808 & 0.00 & circular, contact \\ 
 Gaia DR3 6377601164079244544 & TIC 273550444 & 23 30 39.07 & -76 05 37.07 & 0.406950 & 0.4823 & \cite{Gaia}                & 3.528069 &      & incorrect Gaia period \\ 
 Gaia DR3 6658753777723561088 & TIC 424913817 & 19 30 35.58 & -49 53 38.16 & 0.383308 & 0.5166 & \cite{Gaia}                & 3.14314  &      & incorrect Gaia period \\ 
 Gaia DR3 6710310913044118912 & TIC 397165192 & 18 40 25.61 & -42 44 41.29 & 0.229518 & 0.6660 & \cite{Gaia}                & 4.49258  &      & incorrect Gaia period \\ 
 Gaia DR3 6721515172747931904 & TIC 89048943  & 18 19 32.89 & -42 43 58.75 & 0.813750 & 0.3386 & \cite{Gaia}                & 0.81357  & 0.00 & circular, contact \\ 
 Gaia DR3 764197934635524224  & TIC 17971482  & 11 23 25.33 & +39 14 54.72 & 0.977353 & 0.5001 & \cite{Gaia}                & 4.69045  &      & incorrect Gaia period \\ 
 Gaia DR3 108826366678468224  & TIC 113804557 & 03 09 39.49 & +21 45 32.62 & 0.520687 & 0.2212 & \cite{Gaia}                & 0.52068  & 0.00 & circular, contact \\  
 Gaia DR3 118026289706004736  & TIC 29018151  & 03 23 08.24 & +27 09 23.00 & 0.566686 & 0.4792 & \cite{Gaia}                & 0.566683 & 0.00 & detached, spot \\  
 Gaia DR3 119237539202927232  & TIC 29001158  & 03 22 11.32 & +29 12 38.85 & 0.257846 & 0.4290 & \cite{Gaia}                & 3.333094 &      & incorrect Gaia period \\ 
 Gaia DR3 1333318690910714112 & TIC 9594952   & 17 17 47.01 & +31 36 01.87 & 0.447337 & 0.4829 & \cite{Gaia}                & 0.447329 & 0.00 & circular, contact \\  
 Gaia DR3 141039201218726528  & TIC 73691548  & 02 40 00.38 & +35 18 16.98 & 0.231691 & 0.5034 & \cite{Gaia}                & 5.156766 &      & incorrect Gaia period \\ 
 Gaia DR3 1495170788248851968 & TIC 158108010 & 14 35 22.41 & +47 43 08.20 & 0.491052 & 0.5918 & \cite{Gaia}                & 0.49105  & 0.00 & circular, contact \\  
 Gaia DR3 1562161832705736576 & TIC 160035956 & 13 35 54.71 & +55 27 41.20 & 0.224107 & 0.3940 & \cite{Gaia}                & 8.248437 &      & incorrect Gaia period \\ 
 Gaia DR3 1594473833746069888 & TIC 235559535 & 15 30 28.18 & +49 42 59.47 & 0.876499 & 0.6502 & \cite{Gaia}                & 0.87651  & 0.00 & circular, contact \\ 
 Gaia DR3 1670307043338198144 & TIC 233183004 & 14 36 34.64 & +67 06 46.50 & 0.369582 & 0.7805 & \cite{Gaia}                & 3.381561 &      & incorrect Gaia period \\  
 Gaia DR3 1860986587518562176 & TIC 230619419 & 20 32 01.93 & +30 22 03.86 & 0.948825 & 0.1402 & \cite{Gaia}                & 0.948825 & 0.00 & detached \\  
 Gaia DR3 1879960786652126080 & TIC 20574958  & 22 20 51.25 & +26 19 33.12 & 0.635391 & 0.3344 & \cite{Gaia}                & 0.635388 & 0.00 & circular, contact \\ 
 Gaia DR3 194161972374950784  & TIC 143957468 & 05 24 59.07 & +41 00 16.90 & 0.548736 & 0.4849 & \cite{Gaia}                & 0.548742 & 0.00 & circular, contact \\ 
 Gaia DR3 1950504013882238592 & TIC 160644838 & 21 32 28.97 & +34 33 35.73 & 0.735529 & 0.1219 & \cite{Gaia}                & 0.735531 & 0.00 & circular, contact \\  
 Gaia DR3 1988664351633304320 & TIC 66355834  & 22 44 57.13 & +49 39 27.28 & 0.461419 & 1.1252 & \cite{Gaia}                & 10.36921 &      & incorrect Gaia period \\ 
 Gaia DR3 2013679340674718208 & TIC 315287999 & 23 10 02.57 & +59 12 06.07 & 0.875185 & 0.8130 & \cite{Gaia}                & 1.750470 &      & incorrect Gaia period \\  
 Gaia DR3 2045382590257000832 & TIC 70728658  & 19 40 50.76 & +32 41 54.96 & 0.797011 & 0.9027 & \cite{Gaia}                & 0.797012 & 0.00 & circular, contact \\ 
 Gaia DR3 2049190473878382848 & TIC 392874750 & 19 20 50.41 & +33 39 07.03 & 0.570197 & 0.3302 & \cite{Gaia}                & 0.570198 & 0.00 & circular, contact \\ 
 Gaia DR3 2102608317862904192 & TIC 158431889 & 19 09 33.95 & +43 05 55.66 & 0.572523 & 0.0816 & \cite{Gaia}                & 0.572525 & 0.00 & detached, circular \\  
 Gaia DR3 2102718445127237504 & TIC 158921487 & 19 15 25.74 & +42 48 45.42 & 0.281165 & 0.3107 & \cite{Gaia}                & 0.281164 & 0.00 & circular, contact \\  
 \hline \end{tabular} }
\end{table*}

\begin{table*}
  \caption{Summarization of the systems, continuation.}  \label{systemsInfo2}
  \scalebox{0.77}{
  \begin{tabular}{c c c c| c c c | c c c }\\[-3mm]
\hline \hline\\[-3mm]
    \multicolumn{4}{c|}{Identification}&   \multicolumn{3}{c|}{Original information}&  \multicolumn{3}{c}{Our verdict}    \\
 Target name                           &  TESS number  & RA [J2000.0]& DE [J2000.0] &  P [d]   &    e   &    Reference               &  P [d]   &  e  & comment/remark    \\  \hline\noalign{\smallskip}
 Gaia DR3 2178622474442049920 & TIC 315580490 & 21 23 06.26 & +56 05 42.60 & 0.512635 & 0.6322 & \cite{Gaia}                & 2.60289  &      & incorrect Gaia period \\ 
 Gaia DR3 2232997481954044800 & TIC 428956975 & 22 59 59.19 & +75 21 41.13 & 0.273621 & 0.7049 & \cite{Gaia}                & 3.62785  &      & incorrect Gaia period \\ 
 Gaia DR3 2292031738716349696 & TIC 264007205 & 20 24 12.80 & +80 13 25.70 & 0.975666 & 0.5344 & \cite{Gaia}                & 1.951330 &      & double period \\  
 Gaia DR3 2330974237252306176 & TIC 254287132 & 23 34 15.38 & -28 42 43.82 & 0.648970 & 0.3326 & \cite{Gaia}                & 0.648978 & 0.00 & circular, detached \\ 
 Gaia DR3 23579890146743552   & TIC 381077027 & 02 30 06.71 & +10 05 11.55 & 0.418637 & 0.3970 & \cite{Gaia}                & 0.418633 & 0.00 & circular, contact \\ 
 Gaia DR3 248660949834366848  & TIC 428461757 & 03 43 17.41 & +48 35 20.60 & 0.449186 & 0.9756 & \cite{Gaia}                & 19.7519  &      & incorrect Gaia period \\ 
 Gaia DR3 2679922295485110016 & TIC 241257847 & 21 58 56.94 & -00 40 39.15 & 0.707605 & 0.1337 & \cite{Gaia}                & 0.707605 & 0.00 & asymmetry, spots \\   
 Gaia DR3 2695456088387520128 & TIC 418380927 & 21 58 14.28 & +03 02 06.16 & 0.964878 & 0.6550 & \cite{Gaia}                & 2.72922  &      & incorrect Gaia period \\ 
 Gaia DR3 278374285202188928  & TIC 356385769 & 04 36 23.17 & +58 02 05.79 & 0.205914 & 0.9190 & \cite{Gaia}                & 11.49195 &      & incorrect Gaia period \\ 
 Gaia DR3 2888711169825554688 & TIC 100682885 & 05 50 29.51 & -34 24 11.30 & 0.979144 & 0.4498 & \cite{Gaia}                & 1.958291 &      & incorrect Gaia period \\ 
 Gaia DR3 289944136822295552  & TIC 381318812 & 01 29 36.12 & +22 25 26.88 & 0.527980 & 0.5661 & \cite{Gaia}                & 0.527980 & 0.00 & spot \\ 
 Gaia DR3 2929627154295066112 & TIC 5123582   & 07 20 07.79 & -21 39 12.99 & 0.819588 & 0.8905 & \cite{Gaia}                & 6.556079 &      & incorrect Gaia period \\ 
 Gaia DR3 3031458977763027968 & TIC 409258839 & 07 18 45.38 & -14 48 44.44 & 0.416498 & 1.0154 & \cite{Gaia}                & 3.600295 &      & incorrect Gaia period \\ 
 Gaia DR3 3055046079154617728 & TIC 187852705 & 07 24 50.00 & -06 36 46.95 & 0.989465 & 0.3884 & \cite{Gaia}                & 10.01178 &      & incorrect Gaia period \\ 
 Gaia DR3 3110645358514521728 & TIC 318049261 & 07 23 07.35 & +00 10 50.66 & 0.670822 & 0.2164 & \cite{Gaia}                & 0.670819 & 0.00 & circular, contact \\ 
 Gaia DR3 3129144706069228928 & TIC 237593901 & 06 52 37.64 & +05 03 30.00 & 0.572676 & 0.4113 & \cite{Gaia}                & 8.458561 &      & incorrect Gaia period \\ 
 Gaia DR3 3185448948476544640 & TIC 55998916  & 04 38 11.83 & -09 07 28.55 & 0.414631 & 0.4752 & \cite{Gaia}                & 13.07111 &      & incorrect Gaia period \\ 
 Gaia DR3 3342499719056832640 & TIC 151470826 & 06 07 42.87 & +12 17 34.10 & 0.247152 & 0.6200 & \cite{Gaia}                & 1.524865 &      & incorrect Gaia period \\ 
 Gaia DR3 3354188455373504896 & TIC 387146111 & 06 50 54.51 & +14 14 04.64 & 0.243156 & 0.4094 & \cite{Gaia}                & 4.072190 &      & incorrect Gaia period \\ 
 Gaia DR3 3357997193715943040 & TIC 387143009 & 06 50 55.78 & +15 47 52.59 & 0.721982 & 0.1881 & \cite{Gaia}                & 0.721984 & 0.00 & eclipsing+pulsating (oEA) \\ 
 Gaia DR3 3363402534611486848 & TIC 47203185  & 07 10 49.15 & +19 28 30.69 & 0.978466 & 0.3893 & \cite{Gaia}                & 4.11231  &      & incorrect Gaia period \\ 
 Gaia DR3 3367845042624220672 & TIC 436951143 & 07 04 15.68 & +22 30 41.06 & 0.853624 & 0.7158 & \cite{Gaia}                & 2.49633  &      & incorrect Gaia period \\ 
 Gaia DR3 3442220510252682112 & TIC 73735477  & 05 30 09.76 & +27 22 22.17 & 0.590224 & 0.8152 & \cite{Gaia}                & 2.23601  &      & incorrect Gaia period \\ 
 Gaia DR3 3451442217354511744 & TIC 312225199 & 05 54 04.81 & +34 01 50.19 & 0.818216 & 0.5932 & \cite{Gaia}                & 0.818214 & 0.00 & detached, circular \\ 
 Gaia DR3 3505446307202626304 & TIC 408552341 & 13 00 34.17 & -21 17 35.76 & 0.648944 & 0.5224 & \cite{Gaia}                & 0.648949 & 0.00 & circular, contact \\ 
 Gaia DR3 353593494861867008  & TIC 293184564 & 02 23 55.86 & +45 54 04.66 & 0.658279 & 0.4338 & \cite{Gaia}                & 1.316531 &      & double period \\ 
 Gaia DR3 3633914658836721664 & TIC 61070821  & 13 40 17.31 & -04 22 43.30 & 0.562120 & 0.3753 & \cite{Gaia}                & 0.562121 & 0.00 & circular, contact \\ 
 Gaia DR3 3654517784458899968 & TIC 446434165 & 14 22 30.70 & +01 27 58.97 & 0.812745 & 0.4887 & \cite{Gaia}                & 0.81272  & 0.00 & circular, contact \\ 
 Gaia DR3 3686703930072434176 & TIC 66715008  & 13 17 42.81 & -00 33 44.45 & 0.398513 & 0.4980 & \cite{Gaia}                & 0.398512 & 0.00 & circular, contact \\ 
 Gaia DR3 4124142393505094784 & TIC 418158253 & 17 42 54.01 & -16 08 12.57 & 0.678674 & 0.4830 & \cite{Gaia}                & 1.35734  &      &  double period \\ 
 Gaia DR3 4146343594854786688 & TIC 113959840 & 18 17 56.35 & -14 36 47.16 & 0.525879 & 0.4827 & \cite{Gaia}                & 1.31125  &      & incorrect Gaia period \\ 
 Gaia DR3 4150425703577596288 & TIC 209769501 & 17 55 47.78 & -13 21 54.36 & 0.689569 & 0.7571 & \cite{Gaia}                & 9.97918  &      & incorrect Gaia period \\ 
 Gaia DR3 4171632366602480256 & TIC 8849198   & 17 56 52.46 & -07 18 36.28 & 0.894208 & 0.7433 & \cite{Gaia}                & 3.82826  &      & incorrect Gaia period \\ 
 Gaia DR3 4205393588382859392 & TIC 122740282 & 19 05 14.19 & -07 24 37.65 & 0.249473 & 0.5197 & \cite{Gaia}                & 0.24947  & 0.00 & circular, contact \\ 
 Gaia DR3 4216146640300354048 & TIC 156588652 & 20 22 13.35 & -07 21 03.59 & 0.853687 & 0.9086 & \cite{Gaia}                & 2.139205 &      & incorrect Gaia period \\ 
 Gaia DR3 4225840798100866048 & TIC 212772286 & 20 29 59.51 & -02 20 40.87 & 0.416796 & 0.1867 & \cite{Gaia}                & 0.416792 & 0.00 & circular, contact \\ 
 Gaia DR3 4232567331064146816 & TIC 375573281 & 20 28 59.39 & +02 11 03.02 & 0.924426 & 0.5784 & \cite{Gaia}                & 0.924402 & 0.00 & circular, contact \\ 
 Gaia DR3 4356935500761084416 & TIC 414498939 & 16 20 46.46 & -03 19 57.54 & 0.929090 & 0.5459 & \cite{Gaia}                & 0.929027 & 0.00 & no TESS$^{\star}$\\ 
 Gaia DR3 4381662001144188544 & TIC 297363952 & 17 06 11.21 & +01 09 49.39 & 0.314469 & 0.1421 & \cite{Gaia}                & 0.314464 & 0.00 & circular, contact \\ 
 Gaia DR3 438445487238653440  & TIC 302073092 & 02 43 06.83 & +48 59 02.21 & 0.846288 & 0.2692 & \cite{Gaia}                & 0.846305 & 0.00 & detached, spot \\ 
 Gaia DR3 438801419768593024  & TIC 428122301 & 02 50 38.52 & +50 12 39.50 & 0.612646 & 0.3859 & \cite{Gaia}                & 0.612640 & 0.00 & detached, evolving spot \\ 
 Gaia DR3 4390698204317220352 & TIC 302434394 & 17 28 46.68 & +06 07 09.72 & 0.804916 & 0.3709 & \cite{Gaia}                & 1.609799 &      & double period \\ 
 Gaia DR3 4455992049496820992 & TIC 354746529 & 15 57 17.82 & +10 57 37.09 & 0.368268 & 0.6782 & \cite{Gaia}                & 2.964295 &      & incorrect Gaia period \\  
 Gaia DR3 4503474871461121792 & TIC 310926769 & 18 00 54.98 & +18 34 39.04 & 0.583118 & 0.1551 & \cite{Gaia}                & 0.583116 & 0.00 & circular, contact \\ 
 Gaia DR3 4524651705941314432 & TIC 346467492 & 18 41 56.15 & +19 27 55.66 & 0.241151 & 0.4038 & \cite{Gaia}                & 3.39728  &      & incorrect Gaia period \\ 
 UCAC4 818-024624             & TIC 237290473 & 20 03 23.19 & +73 31 45.80 & 0.83244  & 0.006  & \cite{TANG2025} & 0.832404 & 0.00 & detached, evolving spots \\ 
 UCAC4 368-053333             & TIC 1190662   & 09 04 09.60 & -16 29 36.74 & 0.88770  & 0.190  & \cite{TANG2025} & 0.88816  & 0.21 & flares, pulsations\\ 
 UCAC4 086-008650             & TIC 140661916 & 04 50 14.87 & -72 49 32.47 & 0.52190  & 0.007  & \cite{TANG2025} & 0.52198  & 0.00 & detached, evolving spot \\ 
 ZTF J175114.65+624700.0      & TIC 219819617 & 17 51 14.69 & +62 47 00.06 & 0.53424  & 0.049  & \cite{TANG2025} & 0.534193 & 0.00 & detached, evolving spot \\ 
 TYC 4534-519-1               & TIC 141525324 & 06 48 59.22 & +79 57 08.35 & 0.72492  & 0.185  & \cite{TANG2025} & 0.724899 & 0.00 & detached, evolving spot \\ 
 TYC 3654-2005-1              & TIC 202601061 & 00 29 35.85 & +52 34 07.43 & 0.66104  & 0.056  & \cite{TANG2025} & 0.660943 & 0.00 & detached, evolving spot \\ 
 ASASSN-V J194855.95+644026.8 & TIC 229510628 & 19 48 56.01 & +64 40 27.09 & 0.52983  & 0.046  & \cite{TANG2025} & 0.529856 & 0.00 & detached, evolving spot \\ 
 CD-44 880                    & TIC 146688260 & 02 52 54.05 & -43 31 20.77 & 0.50307  & 0.040  & \cite{TANG2025} & 0.503060 & 0.00 & detached, evolving spot \\ 
 CD-43 14876                  & TIC 279254042 & 22 05 33.41 & -43 20 43.84 & 0.59685  & 0.026  & \cite{TANG2025} & 0.596874 & 0.00 & detached, evolving spot \\ 
 UCAC4 464-001419             & TIC 344715746 & 01 02 49.63 & +02 44 03.83 & 0.938    & 0.02   & \cite{Guo2025} &    --    &      & not eclipsing, pulsating star  \\
 UCAC4 662-053425             & TIC 56913729  & 09 17 42.91 & +42 13 32.31 & 0.924    & 0.03   & \cite{Guo2025} & 0.92397  & 0.00 & detached, spot  \\
 LY Lyn                       & TIC 17859763  & 08 01 51.52 & +41 32 35.46 & 0.851       & 0.03   & \cite{Guo2025} & 0.459314 & 0.00 & incorrect LAMOST period, spot \\ 
 UCAC4 570-022931             & TIC 429432073 & 06 00 55.72 & +23 48 02.27 & 0.823    & 0.02   & \cite{Guo2025} & 4.6949   &      & incorrect LAMOST period \\ 
 AP LMi                       & TIC 16876554  & 10 35 36.23 & +37 46 37.87 & 0.779    & 0.07   & \cite{Guo2025} & 3.5679   &      & incorrect LAMOST period \\ 
 Gaia DR3 206981418960688128  & TIC 410384667 & 04 59 51.26 & +46 37 49.98 & 0.591    & 0.03   & \cite{Guo2025} & 1.44939  &      & incorrect LAMOST period \\ 
 SX Gem                       & TIC 46778677  & 06 28 14.41 & +20 33 50.01 & 0.577       & 0.08   & \cite{Guo2025} & 1.366875 &      & incorrect LAMOST period \\
 1SWASP J221621.13+293307.4   & TIC 28061267  & 22 16 21.13 & +29 33 07.18 & 0.539       & 0.02   & \cite{Guo2025} & 0.350261 & 0.00 & incorrect LAMOST period \\ 
 ASASSN-V J113907.52+343154.1 & TIC 359007937 & 11 39 07.51 & +34 31 54.26 & 0.352    & 0.04   & \cite{Guo2025} & 2.975585 &      & incorrect LAMOST period \\ 
    \hline
\end{tabular}
}
 {\small $^{\star}$ - The only system having no TESS data available. }
\end{table*}

\section{Why a 1~day limit?}

Why was exactly the 1-day period limit chosen? There are two main reasons for this value.

At first, \cite{2006ApJ...651.1151A} shows that for massive B-type stars their circularization
period is about 1.4~days, being significantly shorter than for the later-type stars having this
value of several days. Moreover, \cite{1977A&A....57..383Z} predicted that such a period would be
even shorter than 1~day.

At second, there is still no known eccentric eclipsing binary having a significantly shorter
orbital period below this value (see below in Section \ref{concl}). Therefore, investigating a
larger number of systems just below one day and testing their eccentricity is well-substantiated.

\section{The data}

Thanks to the huge databases provided by Gaia and TESS, a few studies were published quite
recently in which the authors also presented systems with short orbital periods (around or below
1~day) and provided their eccentric orbits. There are especially these papers:
\begin{itemize}
    \item \cite{TANG2025} - database of eccentric eclipsing binaries from TESS after 2~years of mission,
    \item \cite{Oddo2026} - catalogue of eclipsing low-mass binaries from the TESS satellite,
    \item \cite{Gaia} - the eclipsing binary catalogue from Gaia DR3,
    \item \cite{Guo2025} - a few eclipsing systems among the SB catalogue of 665 binaries from the LAMOST Survey.

\end{itemize}

Only as a remark on the Gaia DR3 catalogue systems: we selected only these systems from
\cite{Gaia}, where the periods are below 1~day, are bright enough ($<$ 13 mag in $G$ filter) and
their eccentricity values are adequately precise (ratio of $\delta e / e< 35 \%$). This data set
comprises about 100 systems, which are given in Table \ref{systemsInfo}.

Based on our experience, we feel that many of the binary systems presented do not exhibit
eccentric orbits. This is exactly the point we would like to prove or provide more robust testing
to determine whether these results are adequately reliable. All systems from the above mentioned
papers were collected and tested.

The compilation of the systems under our analysis is given in Table~\ref{systemsInfo}. The
following are provided: the identification of the particular star, its position in the sky, and
information about its inferred period and eccentricity as taken from the specific publication (see
above). Then these periods and eccentricity values are compared with our derived values.

\begin{table*}
  \caption{Incorrectly identified systems.}  \label{IncorrectIdent}
    \scalebox{0.78}{
  \begin{tabular}{c c c c| c c c c| c }\\[-3mm]
\hline \hline\\[-3mm]
    \multicolumn{4}{c|}{Original identification}                                    &   \multicolumn{4}{c|}{Correct identification}   & \\
 Target name                           &  TESS number  & RA [J2000.0]& DE [J2000.0] & Target name                  &  TESS number  & RA [J2000.0] & DE [J2000.0] & Separ. [$^{\prime\prime}$]\\  \hline \noalign{\smallskip}
 Gaia DR3 535964265885838720  & TIC 351611000 & 00 52 28.53 & +69 53 05.37 & ZTF J005227.76+695255.6      & TIC 351611011 & 00 52 27.77 & +69 52 55.61 & 10.5 \\
 Gaia DR3 452750099161604736  & TIC 302233773 & 02 45 00.92 & +52 39 21.65 & ZTF J024501.03+523926.0      & TIC 302233777 & 02 45 01.05 & +52 39 25.93 & 4.4  \\
 CD-44 880                    & TIC 146688260 & 02 52 54.05 & -43 31 20.77 & CRTS J025252.0-433112        & TIC 146688261 & 02 52 52.07 & -43 31 12.26 & 23.2 \\
 Gaia DR3 3000683922169783296 & TIC 34311799  & 06 19 27.31 & -12 22 39.28 & UCAC4 389-012987             & TIC 34311803  & 06 19 28.03 & -12 22 28.80 & 15.2 \\
 Gaia DR3 5376567883659319040 & TIC 162207682 & 11 15 07.30 & -44 46 48.09 & CRTS J111507.6-444702        & TIC 162207678 & 11 15 07.51 & -44 47 02.43 & 14.5 \\
 Gaia DR3 5860580289609692032 & TIC 309679390 & 12 19 14.10 & -65 57 28.51 & Gaia DR3 5860580495751059072 & TIC 309679409 & 12 19 16.66 & -65 57 12.43 & 22.4 \\
 Gaia DR3 4523378711985803392 & TIC 157669515 & 18 20 53.12 & +17 27 21.75 & Gaia DR3 4523379433540312704 & TIC 157669519 & 18 20 51.93 & +17 27 22.93 & 17.0 \\
 Gaia DR3 2198042564210216448 & TIC 469674899 & 21 57 03.94 & +55 13 10.23 & Kiso J215703.51+551314.0     & TIC 469674893 & 21 57 03.52 & +55 13 14.12 & 5.3 \\
  \hline
\end{tabular}
 }
\end{table*}

\begin{figure}
    \centering
      \includegraphics[width=0.88\linewidth]{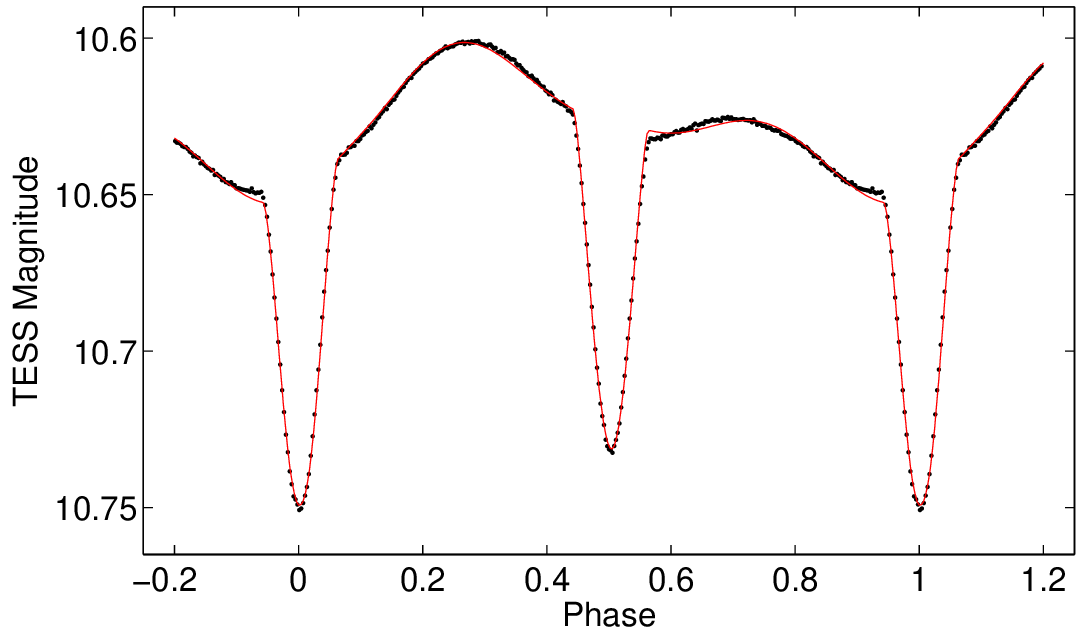}
      \includegraphics[width=0.95\linewidth]{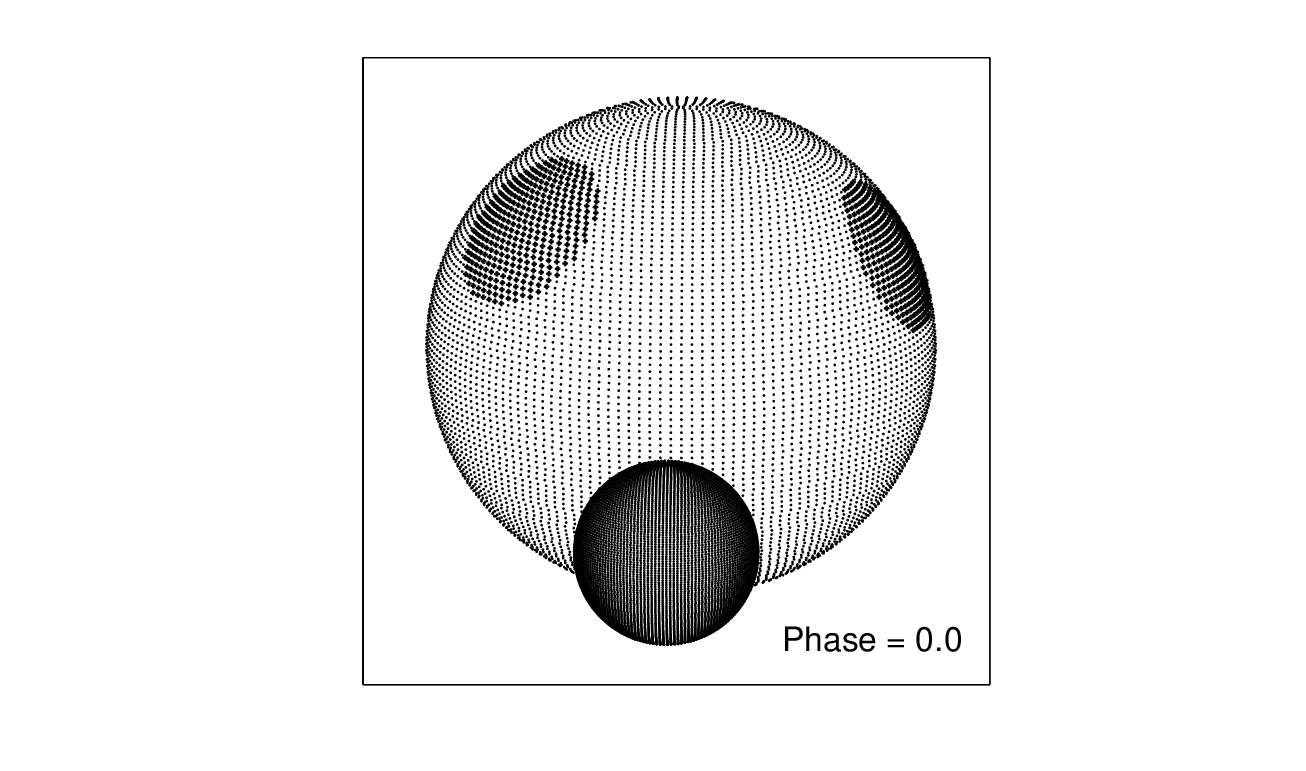}
    \caption{An example of a false eccentric orbit. {\sc Phoebe} light curve solution and 3D model of TIC~279254042. Detached configuration with two dark spots on the primary component makes the automatic analysis for pipelines problematic, producing sometimes spurious eccentricities.}
    \label{fig:example}
\end{figure}

\begin{figure}
  \centering
   \includegraphics[width=0.45\textwidth]{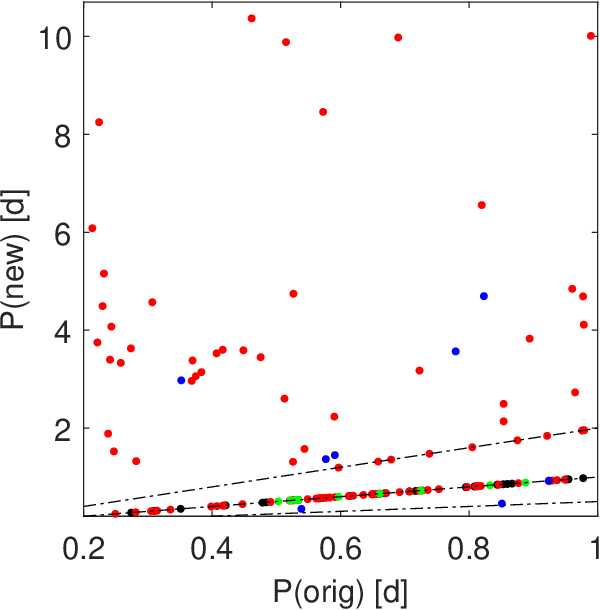}
 \caption{Period-period diagram showing the original values of period, as taken from different
publications, compared with our new derived correct value of period. The 1:1 ratio is also shown
as dash-dotted line (corresponding to the correct value), together with two other lines showing
half and double value of period. Different sources of data are shown with different colours: black
denote Oddo et al.(2026), red denote Mowlavi et al.(2023), green denote Tang et al.(2025), and blue denote
Guo et al.(2025).}    \label{p-p_diagram}
 \end{figure}

\begin{figure}
  \centering
   \includegraphics[width=0.49\textwidth]{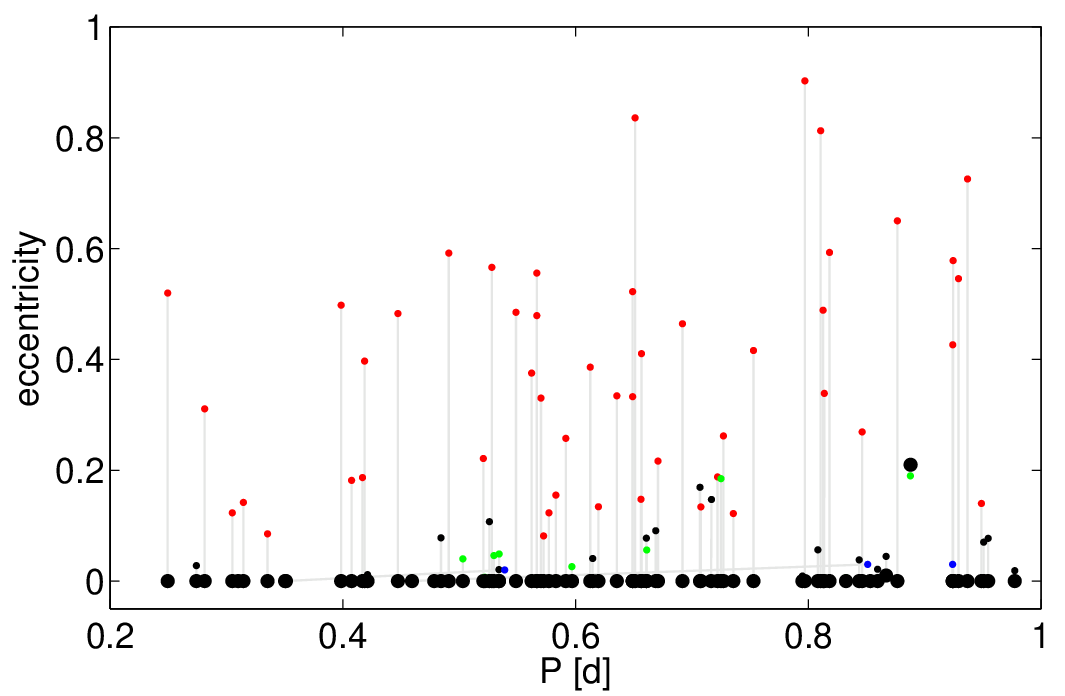}
   \caption{Period-eccentricity diagram comparing the original eccentricities with our new derived ones. The original and new ones are connected with the grey lines, otherwise the colour notation is the same as in Fig.~\ref{p-p_diagram}.}
   \label{P-e_diagram}
 \end{figure}

\section{Methods used and results}

Our approach for the whole analysis was to use only the eclipsing binaries for modelling.
Moreover, such systems need to be short-periodic enough. Owing to this reason, the use of the TESS
photometric data \citep{2015JATIS...1a4003R} is crucial for the whole analysis here. The
photometric data points for the particular system were extracted from the TESS data archive using
the freely-available code named Lightkurve, which is a Python package for data analysis
\citep{2018ascl.soft12013L}. This software is able to reduce the rough data pixel-by-pixel and to
provide us with the flux values on that pixel in every frame recorded. For different stars,
different aperture sizes were chosen (most typically 3x3 pixels). Before the light curve solution
usually some (lower-order) polynomial fitting of these data need to be done to subtract the
various trends in one TESS sector. The light curve solution was then done on the best TESS sector
available (usually with the highest cadence of points, i.e. with the highest number of data
points).

Then, the light curves were analysed using the software {\sc{PHOEBE}} \citep{2005ApJ...628..426P},
which is originally based on the  Wilson-Devinney algorithm, see \cite{1971ApJ...166..605W}. This
programme was used to testing whether the particular binary is eccentric or circular. The use of
eclipsing binaries is a logical step because they can provide fresh insight into the periods and
eccentricities directly: due to dense photometry from the TESS satellite, one can derive the
orbital period easily, and to obtain eccentricity value is also quite straightforward because of
modelling the light curves (eclipse duration versus phase separation of eclipses, no matter in
which phase of the apsidal motion).

For the whole modelling, we have to use several assumptions. This is mainly due to the unavailable
data for detailed analysis, especially the missing spectroscopy. Having only the photometry in our
hands, we have to fix the mass ratio to unity ($q$ = 1.0) for detached systems with negligible
ellipsoidal out-of-eclipse variations (see e.g. \citealt{2005Ap&SS.296..221T}). The relative radii
$R_i/a$, inclination $i$, and relative temperatures $T_1/T_2$ were obviously the standard free
parameters to be converged. However, for the detailed modelling of the light curve, when the
superb TESS quality of the data allows us to do so, we also let converge the parameters like the
albedo coefficients $A_i$, limb-darkening $x_i$, and gravity brightening $g_i$. In addition, the
third light level was set as a free parameter. The fixed value not fitted was the period, which
was set from all available TESS sectors, best fitting all of them. The synchronicity parameters
$F_i$ were fixed at 1.0 for all of the systems.

\subsection{Properties of the whole population}

From the total number of 146 systems in the input, our results show that 142 are true eclipsing
binaries, albeit with significantly different periods. This is especially true for the subsample
of stars taken from the publication of \cite{Gaia}, i.e. the Gaia DR3 eclipsing binaries
compilation. Such work suffers from incorrect period derivation, mostly due to the lack of data
points. This result can be clearly seen in our Fig.\ref{p-p_diagram}, where we plotted the
original periods and our derived correct values of the periods. As can be seen there, the Gaia~DR3
binaries have mainly incorrect periods, and their true values are usually longer, sometimes even
exceeding the 10-day period. For the particular binary, see  Table~\ref{systemsInfo}. The overall
statistics of the Gaia binaries shows that from the total number of 104 systems, 49~were found to
have incorrect periods (and moreover, for all of them their newly derived periods are always
longer than the 1-day period limit), four others have double values of periods compared to Gaia,
and the rest have periods below the 1-day period limit, but none of them show eccentric orbits.
Therefore, one should be very cautious when using the Gaia derived periods and/or eccentricities
as originally published by \cite{Gaia}.

The most problematic issue was definitely the role of the spots and their time evolution. Usually,
the values of eccentricities are derived using the standard method of measuring the eclipse
durations and eclipse separations in the phased light curve plots. These values can provide us
with information about the $e \cos(\omega)$, and $e \sin(\omega)$, which can later be transformed
into eccentricity values. However, such an approach can be rather tricky when used blindly by an
automated pipeline or algorithm that is not properly validated with some human supervision. See
our Fig.~\ref{fig:example} as an example of such a light curve. Many spurious values can be seen
in the literature over the last few years due to the use of automated algorithms.

Besides the spots and their evolution, a non-negligible group of stars showing flares was also
found. This is also quite expected since we mostly deal here with cool late-type stars, where
flares are quite common phenomena.

On the other hand, apart from the spots and flares, the most problematic issue found among the
binaries from \cite{Oddo2026} was the incorrect identification of the star showing eclipses. This
is obviously due to the large TESS pixels and poor angular resolution. We provide additional
information about the true source of the periodicity below in Table \ref{IncorrectIdent}. As one
can see, sometimes the true source of variability is more than 20$^{\prime\prime}$ away from the
originally assumed one.

For systems below the period limit of 1 day, their light curves were solved using {\sc{PHOEBE}}
and the eccentricity was tested as a free parameter. However, for the vast majority of the
systems, their orbits resulted in a perfectly circular shape. For the systems, we used an inferred
detection threshold of 0.003 in eccentricity. This detection threshold of 0.003 in eccentricity
was set due to the typical uncertainties of the eccentricity values derived during the modelling.
For the best available TESS sectors the derived eccentricity was able to be even at the level of
0.001, but this is mostly not the case. The problem is usually with the spots, and their evolution
in time, what makes better derivation of eccentricity impossible. Other studies were using
typically 0.01 as a limiting value for $e$, or $e \cos \omega$
\citep{2024A&A...691A.242I,{2005ApJ...620..970M}}. Except for the two systems (TIC~1190662, and
TIC~157669515, which is in fact TIC~157669519) all other stars move on perfectly circular orbits.
A probable reason for flagging these systems as eccentric ones is given in the last column of
Table~\ref{systemsInfo}.

 \begin{table}
\caption{Light curve parameters for TIC 1190662, and TIC~157669519.}
 \label{TabLCfit}
  \centering
\begin{tabular}{c c c}
   \hline\noalign{\smallskip}
                & \multicolumn{1}{c|}{TIC 1190662}   & \multicolumn{1}{c}{TIC~157669519}  \\[0.5mm] \hline
 $HJD_0$ [d]         & 2461052.959 $\pm$ 0.002   & 2460186.2826 $\pm$ 0.001 \\
 $P$ [d]             & 0.8881675 $\pm$ 0.0000016 &  0.8668385 $\pm$ 0.0000002  \\
 $e$                 &    0.21 $\pm$ 0.01        &  0.01 $\pm$ 0.01  \\
 $\omega$ [deg]      &   305.0 $\pm$ 0.8         &    93.2 $\pm$ 1.5 \\
 $i$ [deg]           &   86.72 $\pm$ 0.55        &   73.82 $\pm$ 0.32  \\
 $q = M_2/M_1$   &     1.0 (fixed)           &    1.0 (fixed)    \\
 $T_1$ [K]           &    3500 (fixed)           &    6300 (fixed)   \\
 $T_2$ [K]           &    3217 $\pm$ 76          &    4894 $\pm$ 89  \\
 $R_1/a$             &   0.090 $\pm$ 0.002       & 0.262 $\pm$ 0.00  \\
 $R_2/a$             &   0.065 $\pm$ 0.002       & 0.262 $\pm$ 0.00  \\
 $L_1$ [\%]          &    4.8 $\pm$ 0.3          &  12.5 $\pm$ 0.6   \\
 $L_2$ [\%]          &    1.6 $\pm$ 0.3          &   4.0 $\pm$ 0.4   \\
 $L_3$ [\%]          &   93.6 $\pm$ 1.0          &  83.5 $\pm$ 2.1   \\ \hline
 \noalign{\smallskip}
\end{tabular}
 \\
  \begin{flushleft}
  \footnotesize Note: The uncertainties of individual parameters are taken from {\sc PHOEBE} only, and are usually underestimated.\\
  \end{flushleft}
\end{table}

\subsection{Two eccentric systems found}

We were able to identify only two systems with obviously eccentric orbits. Here we would like to
draw special attention to both of them. For the final confirmation that the orbits are truly
eccentric, one would need to combine different proofs of eccentricity: light curve solution,
eclipse timings, and the radial velocities, if possible. Owing to the fact that both of these
systems are rather faint (GAIA $G > 14$~mag), to obtain their radial velocity curve would be a
rather difficult task. Moreover, with their periods below 1 day, this would only be achievable
with 8-m class telescopes to obtain a sufficient S/N ratio. Therefore, we used only the
photometric methods.

\begin{figure}
  \centering
  \hspace{-8mm}
   \includegraphics[width=0.41\textwidth]{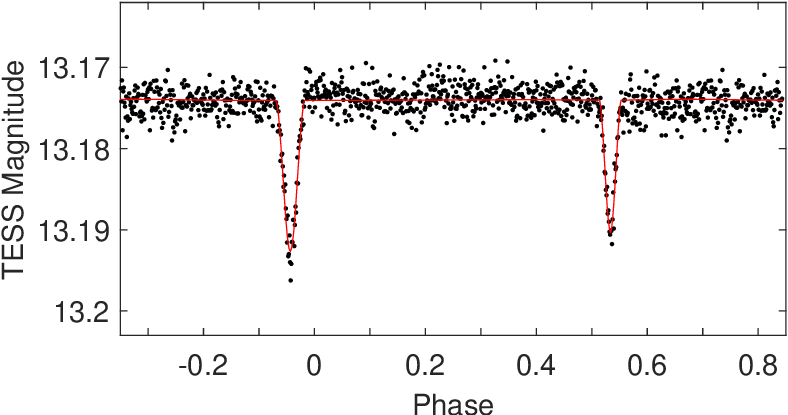}
   \vskip 9mm
   \includegraphics[width=0.46\textwidth]{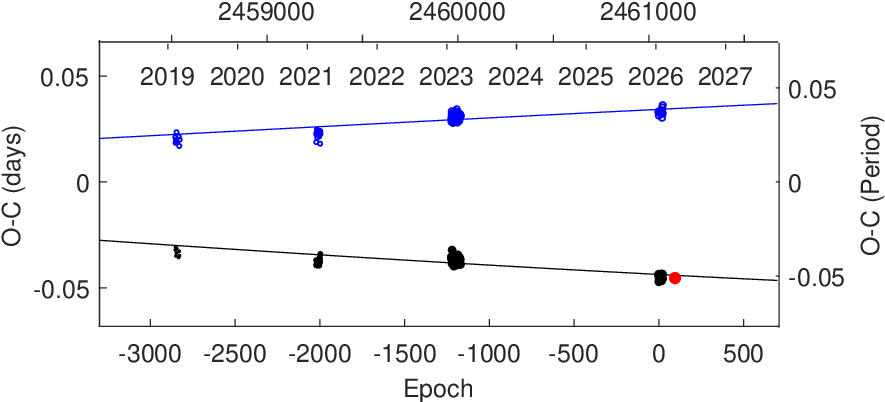}
   \caption{The system TIC 1190662. Upper plot shows the eclipsing light curve from the TESS data (sector 62). Lower plot is the O-C diagram of times of eclipses, blue ones are the secondaries, black ones are the primaries. The red dot is our own observation from the Ond\v{r}ejov observatory, Czech Republic. The diverging curves show the slow apsidal motion. }
   \label{TIC1190662}
 \end{figure}

Let us first look at the system TIC~1190662. A relatively high value of eccentricity was found,
namely 0.21. This value is well-substantiated due to the independent fitting of different TESS
sectors, which yielded very similar values (between 0.210 and 0.212). Such an eccentric orbit can
possibly be induced by some three-body interaction, probably present there (it shows a complicated
light curve shape, flares, second periodicity, maybe some pulsations, and also visible
eclipse-timing variation (ETV)). The curve after subtracting the additional variability is plotted
in Fig.\ref{TIC1190662}, results given in Table \ref{TabLCfit}. There is clearly seen the offset
of both eclipses away from 0.0 and 0.5 phases. On the bottom plot, there is also plotted the slow
apsidal advance of the orbit in space, which is still only poorly covered now, but leads to the
apsidal period of about 170~yr. Unfortunately, due to its very small photometric amplitude and
relatively faint brightness, the star was not identified as an eclipsing one in other databases or
surveys, hence we cannot extend the time base of the data to better constrain the apsidal advance.
The only one data point apart from the TESS is our new observation. Small additional ETV variation
after subtracting the apsidal motion is at the level of 0.005~days, but cannot be better
constrained due to the poor coverage of eclipsing data now. Maybe some future TESS data (planned
to be obtained in 2027) will shed some light to these variations.

The light curve analysis of this system was performed using the assumption of the primary
temperature of 3500~K. This value resulted as the best from the SED fitting (VOSA, see
\citealt{2008A&A...492..277B}), and is also in good agreement with other published values from
different surveys and estimates (like GAIA, TESS, Atlas, etc. ranging from 3300 to 3657~K).

The second eccentric system found is TIC~157669519, which seems to be a more challenging system.
The main problem is its very small eccentricity value, which is only about 0.01. However, this
value independently resulted from several TESS sectors of data (individual fits resulted in values
from 0.009 to 0.014). However, definitively proving an eccentric orbit of such low eccentricity is
somewhat tricky. Older photometric data are not able to show any significant eccentricity, but
also the times of eclipses as derived from the TESS sectors show a scatter larger than the
expected difference between the primary and secondary eclipses on the inferred eccentric orbit.

\begin{figure}
  \centering
  \hspace{-8mm}
   \includegraphics[width=0.41\textwidth]{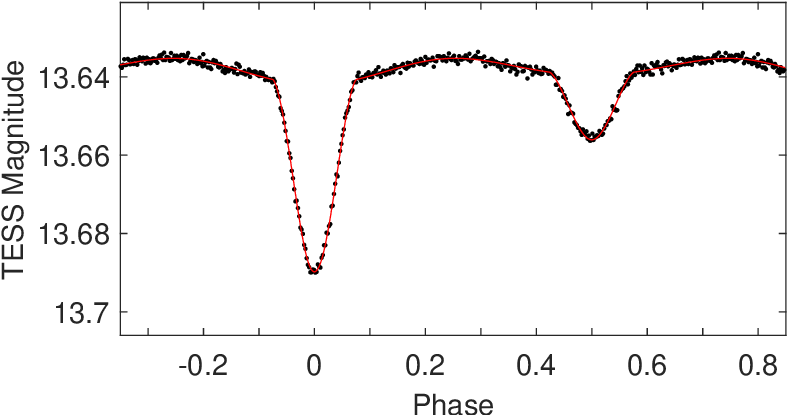}
   \vskip 9mm
   \includegraphics[width=0.46\textwidth]{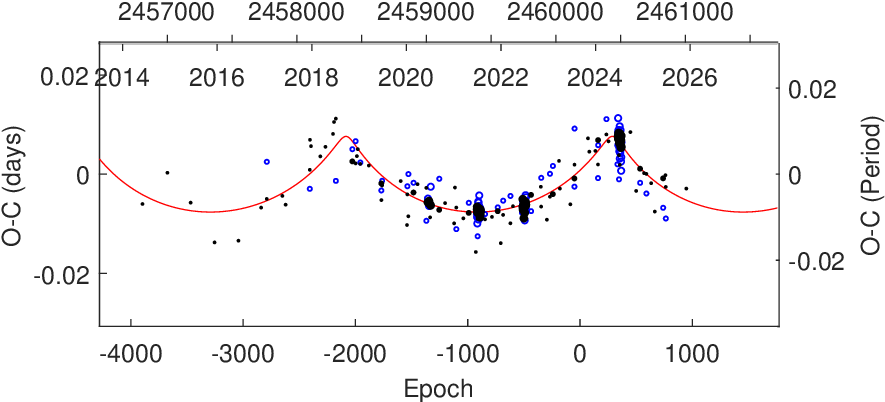}
   \caption{The system TIC 157669519. Upper plot shows the eclipsing light curve from the TESS data (sector 80). Lower plot is the O-C diagram of times of eclipses, blue ones are the secondaries, black ones are the primaries. The red solid curve represents the light-time effect fit to the available data, in agreement with the hypothesis of the third body in the system. }
   \label{TIC157669519}
 \end{figure}

The results of our fitting are given in Fig.\ref{TIC157669519}, where one can see the light curve
fit together with the O-C diagram of the available eclipse times. For the light curve modelling,
the assumption of a 6300~K primary star was used, following the two spectra obtained using the
LAMOST survey \citep{2026yCat.5162....0L}, who classified the star as F7 type. Due to the
relatively deep eclipses, it was possible to derive many eclipsing times from various photometric
surveys apart from TESS, such as Atlas, ASASsn, and ZTF. This dataset spreading over more than 10
years shows a clear period variation. Using a standard hypothesis of the third body orbiting
around an inner eclipsing binary, a so-called light-time effect \citep{1990BAICz..41..231M}, one
gets the following parameters of such a fit: period 5.63 yr, amplitude 0.0076 days, eccentricity
0.73.

\section{Conclusions} \label{concl}

The definitively proved, still eccentric eclipsing system with the shortest period in our Galaxy
was to the best of our knowledge until today, the system V456~Oph \citep{2017AcA....67..257W}. The
systems shows besides fast apsidal motion also some putative third-body variations from its ETV.
Its orbital period is about 1.016 days. On the other hand, there possibly exists one extragalactic
eccentric system below the 1-day period limit, namely the star OGLE LMC-ECL-17226
\citep{2020A&A...640A..33Z}, which has a period of about 0.9879 days. However, besides that we are
not aware of any other proven eclipsing system with a reliable eccentric orbit and an orbital
period shorter than 1~day.

According to the circularization theories (see the references above in Section~1), there is a
difference between the effectivity of the tidal circularization processes in early and late type
stars, see e.g. \cite{2019EAS....82..127S}. Hence, one can ask whether the two binaries currently
detected and presented in our sample agree with this finding. Both of our two detected eccentric
systems are rather late-type stars. Nevertheless, it is necessary to mention also a possibility
that the third-body dynamics can play a role here, since both systems show some ETV in their
minima timings.

Different circularization theories (\citealt{1977A&A....57..383Z}, \citealt{1992ApJ...395..259T},
and \citealt{1997A&A...318..187C}) present various circularization period cut-off values (i.e. the
borderline between circular, and eccentric regimes), being the most probable in between 1 and
10~days. Here we provide for the first time a proof that there definitely exist systems with
eccentric orbits also below the 1-day period limit. Apart of pointing out to many published
spurious orbits, which are in fact circular, we should ask whether our finding has some wider
consequences or theoretical implications. Detecting systems shorter than one day and having in
mind that there definitely exist systems with reliable eccentric orbits just above the 1-day
period, we can safely say that the circularization period definitely cannot be at the level of
10~days. However, do these two examples shift the correct circularization period to the values
below 1 day? Therefore, are the circularization theories incorrect, or badly parametrized? We do
not think so. The reason is the following. Having only two cases is still only weak evidence and
low-number statistics for any reliable conclusions. One of the systems is rather questionable,
having the eccentricity at the level of its error. The second system can have its higher
eccentricity induced by a third-body dynamics. The Kozai-Lidov
\citep{1962AJ.....67..591K,{1962P&SS....9..719L}} cycles are a typical mechanism to push the
eccentricity value of the inner orbits to the higher values due to the influence of the third
body. In our single system, we cannot be sure that such an effect can be ruled out without a
proper and detailed modelling with long term monitoring and spectroscopic data.

Therefore, it is still too premature to say anything about the true shift of the circularization
period here. However, what about the whole population of our systems and their link to the other
groups of stars studied? Unfortunately, our studied eclipsing systems are just random normal field
stars, having only very little in common, not being parts of any significant groups or clusters.
Their ages, compositions, and structure are different. On the other hand, what can be said even
now, after analysing almost 150 systems, and resulting in identification of only one trustworthy
short-periodic eccentric system: only adequately comprehensive analysis of particular system can
bring us undisputable evidence on the orbital eccentricity.

\section*{Acknowledgments}
An anonymous referee is acknowledged for his/her helpful and critical comments, greatly improving
the overall quality of the manuscript. The research was supported by the project {\sc Cooperatio -
Physics} of Charles University in Prague. This research made use of Lightkurve, a Python package
for TESS data analysis \citep{2018ascl.soft12013L}. This research has made use of the SIMBAD and
VIZIER databases, operated at CDS, Strasbourg, France, and of NASA Astrophysics Data System
Bibliographic Services. This work has made use of data from the European Space Agency (ESA)
mission {\it Gaia} (\url{https://www.cosmos.esa.int/Gaia}), processed by the {\it Gaia} Data
Processing and Analysis Consortium (DPAC,
\url{https://www.cosmos.esa.int/web/Gaia/dpac/consortium}). Funding for the DPAC has been provided
by national institutions, in particular the institutions participating in the {\it Gaia}
Multilateral Agreement. This publication makes use of VOSA, developed under the Spanish Virtual
Observatory (https://svo.cab.inta-csic.es) project funded by MCIN/AEI/10.13039/501100011033/
through grant PID2020-112949GB-I00. VOSA has been partially updated by using funding from the
European Union's Horizon 2020 Research and Innovation Programme, under Grant Agreement nº 776403
(EXOPLANETS-A).

\section*{Data availability}

All the data used in the manuscript, which are not already included in the Tables, will be shared
upon a reasonable request to the corresponding author.

\end{document}